\documentclass[prb,preprint]{revtex4}
\usepackage{graphicx}
\usepackage{dcolumn}
\usepackage{bm}
\usepackage{amssymb,amsmath}

\begin{document}
\title{A probabilistic explanation for the size-effect in crystal plasticity}
\author{P. M. Derlet}
\email{Peter.Derlet@psi.ch} 
\affiliation{Condensed Matter Theory Group, Paul Scherrer Institut, CH-5232 Villigen PSI, Switzerland}
\author{R. Maa{\ss}}
\email{robert.maass@ingenieur.de}
\affiliation{Institute for Materials Physics, University of G\"{o}ttingen, Friedrich-Hund-Platz 1, D-37077 G\"{o}ttingen, Germany}
\date{\today}
\begin{abstract}
In this work, the well known power-law relation between strength and sample size, $d^{-n}$, is derived from the knowledge that a dislocation network exhibits scale-free behaviour and the extreme value statistical properties of an arbitrary distribution of critical stresses. This approach yields $n=(\tau+1)/(\alpha+1)$, where $\alpha$ reflects the leading order algebraic exponent of the low stress regime of the critical stress distribution and $\tau$ is the scaling exponent for intermittent plastic strain activity. This quite general derivation supports the experimental observation that the size effect paradigm is applicable to a wide range of materials, differing in crystal structure, internal microstructure and external sample geometry.
\end{abstract}
\maketitle

\section{Introduction} \label{SecIntro}

\begin{figure}
\begin{center}
\includegraphics[clip,width=1.0\textwidth]{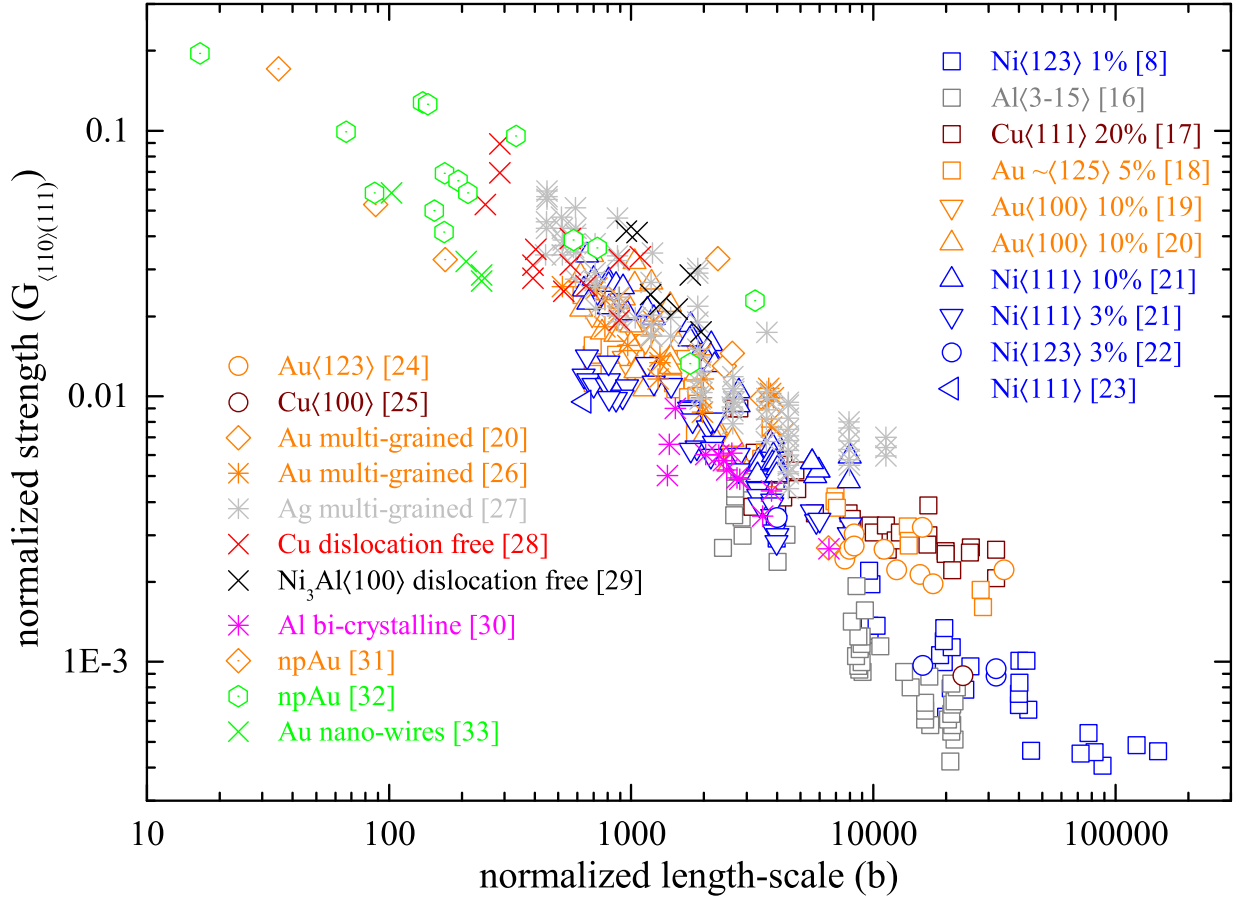}
\end{center}
\caption{Log-Log plot of strength normalized to the appropriate shear modulus versus sample size normalized to the Burger's vector magnitude for a wide range of literature data.} \label{FigExpData} 
\end{figure}

Smaller is stronger. This is the most general conclusion that can be drawn from numerous experimental and theoretical studies investigating the plastic flow behaviour of metallic materials. Examples are the empirical Hall-Petch relationship~\cite{Hall1951,Petch1953}, strain-gradient strengthening \cite{Fleck1997,Zaiser2003,Zhang2011}, indentation size-effects~\cite{Nix1998,Begley1998}, and the most recent observation of a sample size-effect due to the reduction of the external dimensions~\cite{Uchic2004,Dimiduk2005}. Whilst size-affected plastic flow as a result of a finite sample size has been reported sporadically ever since G.F. Taylor's work in 1924~\cite{Taylor1924}, intensely focused research emerged first in the past two decades, primarily motivated by production routes and test systems that allow systematic and well controllable experiments at the micron- and nano-scale.
The central finding in recent developments on finite sample-size effects is an empirical power-law scaling of the type, $\sigma_{\mathrm{flow}}\sim d^{-n}$, with $d$ the characteristic length-scale of the sample and $n$ a power-law exponent. Summarising size-dependent strengths for fcc metals in a plot containing the strength normalised by the shear-modulus and the sample dimension normalised by the Burgers vector yields a surprisingly general trend for all data with a power-law exponent of typically around $n\simeq0.6-0.7$. For bcc crystal systems this power-law scaling holds as well, but here the normalized data exhibits a less universal trend, with $n$ ranging between 0.3 and 0.8, depending on the metal~\cite{Schneider2009}. Fig.~\ref{FigExpData} reproduces some selected data for various fcc metals, clearly demonstrating a quite general strength-size scaling with respect to finite sample size that covers more than two orders of magnitude in both strength and size. A more complete set of data for both fcc and bcc metals can be found in Refs.~\cite{Uchic2009,Greer2011,Kraft2010,Dou2009}. 

Without doubt, the scaling depicted in fig.~\ref{FigExpData} represents a truly remarkable result. How can one explain this general trend for such a variety of experimental studies? In fact, why does the scaling survive the large variations in microstructure, crystal orientation, strain hardening response, testing condition and other strength influencing factors? It is noted that fig.~\ref{FigExpData} contains data from focused ion beam (FIB) prepared single crystals~\cite{Dimiduk2005,Ng2008,Kiener2006,Volkert2006,Greer2006,Greer2005,Frick2008,Maass2012,Shan2008,Maass2009,Maass2009a}, multi-grained electroplated crystals~\cite{Greer2006,Greer2005}, multi-grained crystals prepared by embossing~\cite{Dietiker2011,Buzzi2009}, nominally dislocation free crystals~\cite{Richter2009,Maass2012a}, bi-crystalline FIB prepared crystals \cite{Kunz2011}, nano-porous structures~\cite{Hodge2007,Biener2006}, and also nano-wires \cite{Dou2008}, all of which are expected to contain very different local environments for the operating dislocations. In addition, some of the data contained in fig.~\ref{FigExpData} is highly affected by geometrical strain hardening, because of low aspect ratios and side-wall taper \cite{Zhang2006,Kiener2009}. This hardening is not only reflected by different slopes of the flow curves, but can also be correlated with the formation of dislocation substructures in tapered samples~\cite{Maass2009,Maass2009b}. On the other hand, a rather constant dislocation density at constant sample size is observed during straining in non-tapered geometries in compression and in tension~\cite{Maass2012,Norfleet2008,Mompiou2012}. Despite these differences in micro structural evolution, the strength values, which typically are derived at arbitrary strains between 1\% and 20\%, as selectively indicated, tend to fall similarly onto fig.~\ref{FigExpData}.
 
When studying more carefully individual data sets within fig.~\ref{FigExpData}, it can be shown for some studies that the scaling exponent is dependent on the strain at which the strength is derived~\cite{Frick2008,MaassThesis}. Modelling flow responses of micron sized samples has also evidenced that the scaling exponent is sensitive to the initial underlying dislocation density and structure \cite{Rao2008,Motz2009}, which subsequently was supported by experimental findings \cite{ElAwady2013,Schneider2013}. Yet, all these influences are blurred by the plotted data in fig.~\ref{FigExpData}, which means that fine details in the microstructure are yielding variations in the value of $n$, but the empirical power-law remains the describing functional form irrespective of the micro structural richness covered within fig.~\ref{FigExpData}. 

First explanations for the trend depicted in fig.~\ref{FigExpData} revolved around the scarcity of available dislocation sources and mobile dislocations~\cite{Rao2008,Parthasarathy2007}, as well as the balance between the dislocation escape rate and the dislocation nucleation rate~\cite{Greer2006,Greer2005}. More recently, further understanding has been gained via detailed and specific mechanisms (or change in mechanisms), suggesting that a range of ``non-universal'' explanations underlie the experimental trend seen in fig.~\ref{FigExpData}, as discussed by Kraft and co-workers~\cite{Kraft2010}. Here, the governing dislocation mechanism changes with decreasing sample size from dislocation multiplication in the micron regime, to nucleation controlled plasticity of full (100-1000 nm) and partial dislocations (10-100 nm). This contemporary viewpoint is well motivated by experimental data obtained at all these scales, but still raises the question of how very different underlying effects and mechanisms lead to the very impressive double-logarithmic scaling? Obviously, the data itself suggests one regime without mechanistic transitions. Indeed, it has been argued that the size effect originates from a simple restriction of the available space for dislocation source operation which, although quite general, results in an exponent restricted to unity~\cite{Dunstan2013}. The exception to the lack of a change of mechanism is in the regime of very small sample sizes, where the power-law scaling seems to level off. A reduction in the scaling at very small sizes, corresponding to extremely low defect densities, or even dislocation free systems, can be explained by the relative ease of partial dislocation nucleation as compared to the nucleation of full dislocations or dislocation multiplication processes~\cite{Kraft2010}, but has been shown to arise in micron-sized systems as well~\cite{Uchic2004}. Besides the scarce experimental data in the sub 100 nm regime, several atomistic studies have predicted either the break-down of the ubiquitously observed power-law~\cite{Zhu2008} or a reduction in scaling exponent due to mobile dislocation exhaustion~\cite{Sansoz2011} at the far left end in Fig.1. In the extreme case of fully dislocation free systems, flow stresses are said to depend on the atomic roughness on the surface, yielding a weak intrinsic size-dependence~\cite{Zhu2010,Maass2012a} or no size-scaling at all~\cite{Bei2007} --- a topic which remains to be fully explored. 

With the above at hand, it becomes clear that the trend in fig.~\ref{FigExpData} comprises a wealth of underlying details, that involves a complex convolution of micro structural properties at the detailed level of individual dislocations or even point defects, without mentioning the numerous external experimental factors that have been discussed extensively in the literature~\cite{Maass2009,Zhang2006,Kiener2009,Shade2009}. In terms of detailed structural mechanisms, the discussed size-scaling opens a practically un-explorable multi-dimensional space of parameters that, however, seem not to substantially affect the uniform trend. Such a situation suggests an entirely probabilistic description of plasticity which considers only the statistics of both stress and plastic strain, and how this might change as a function of sample volume. Indeed, Zaiser in his review of intermittent plasticity~\cite{Zaiser2006} proposed that any size dependence will most likely emerge from a change in sampling statistics. 

The above viewpoint has been followed in a number of recent works. For example Demir {\em et al} \cite{Demir2010} have assumed a distribution of source lengths which, when combined with the stress to bow out such a dislocation source, results in a distribution of critical stresses. For the bulk regime all source lengths admitted by the distribution are possible and the critical stress scale is set by the mean value of the distribution. However as the sample volume reduces the source length distribution must be truncated and at a sufficiently small sample length scale the bulk mean field picture breaks down, with the critical stress scale being set by the statistics of small sources and corresponding high critical stresses. On the other hand Pharr and co-workers ~\cite{Phani2013} assumed that the yield strength depends only on the spatial distribution of dislocations and on the distribution of their activation strengths. Despite the lack of specific dislocation mechanisms, an averaging of the resulting yield strength over a certain system size range demonstrates a cross over between a bulk strength at larger system sizes and close to the theoretical strength at very small system sizes. The transition range is the regime of the apparent power-law scaling, which in the work of Phani et al. ~\cite{Phani2013} is uniquely determined by the theoretical strength and the bulk strength, and not by the dislocation density or specimen size.

In the current paper a quite different probabilistic approach will be taken. In particular, using extreme value statistics and an assumed distribution of critical stresses characterized by an algebraic exponent $\alpha$, sec.~\ref{SecStress} derives a very general size effect scaling for stress. Sec.~\ref{SecStrain} then combines this stress scaling with the known scaling exponent, $\tau$, of plastic strain magnitudes for the two regimes of a dominant internal length scale and external length scale. For the case of an internal length scale no size effect emerges. However for the case of an external length scale, a power law in strength emerges where the exponent is given by $n=(\alpha+1)/(\tau+1)$.

\section{Extreme value statistics and a size effect in stress} \label{SecStress}

The applicability of the size effect to a range of materials whose underlying microstructures are expected to be different in their details, motivates the need for a quite general approach that cannot depend too strongly on the specifics of a particular material. One starting point is to acknowledge that bulk plasticity arises from irreversible structural transformations whose core regions generally have a finite spatial extent, and that each leads to a global plastic strain increment which, however small, is discrete. This latter aspect is motivated by the early torsion experiments of Tinder {\em et al} \cite{Tinder1964,Tinder1973} who, with a strain resolution of $10^{-8}$, where able to observe discrete plasticity in bulk metallic samples. Such structural transformations may be characterised, in the first instance, by a critical stress needed for the plastic event to occur with a particular degree of certainty. This may be done by assuming that the model material is defined by a probability distribution of such critical stresses, and that the corresponding number density of irreversible structural transformations is given by the product of this distribution with the total number of distinct structural transformations admitted by the system, $M$. 

The actual distribution of critical stresses will embody the details of the particular material through its underlying low energy potential energy landscape. This latter contribution arises directly from the assumption that plasticity occurs via thermal activation and as a result the distribution will have an implicit strain rate and temperature dependence. In the present work, the aforementioned distribution is assumed to have the form 
\begin{equation}
P[\sigma]=\frac{\delta}{\Gamma\left(\frac{\alpha+1}{\delta}\right)\sigma_{0}}\left(\frac{\sigma}{\sigma_{0}}\right)^{\alpha}\exp\left[-\left(\frac{\sigma}{\sigma_{0}}\right)^{\delta}\right] \label{EqProbDist}
\end{equation}
where $\alpha$, $\delta$ and $\sigma_{0}$ are positive non-zero numbers, and $\Gamma(x)$ is the gamma function. Such a distribution is called the generalized gamma distribution and spans a range of well known positive valued distributions such as the Weibull distribution (and therefore the Rayleigh distribution) with $(\alpha+1)/\delta=1$ and the gamma distribution (and therefore the chi and chi-squared distributions) with $\delta=1$. The sample volume, $V$, enters the model via $M$ since $M$ is an extensive quantity where $M=\rho V$. Here $\rho$ is the density of available irreversible structural excitations.

For a particular realisation of the material system, $M$ critical stress values are sampled from the stress distribution. The applied stress at which the first plastic event occurs will equal the lowest critical stress of these $M$ critical stresses. To proceed further, an assumption has to be made as to how the distribution changes with increasing plastic activity. Presently, it assumed that the analytical form of the distribution and $M$ do not change. These assumptions will be discussed in sec.~\ref{SecDiscussion}. Thus, the next step necessarily involves re-sampling the same distribution to return the system to its $M$ values of critical stress. This re-sampling is performed until a value is found that is greater than the current applied stress implying that, although the form of the intrinsic distribution does not change, the distribution naturally becomes truncated and correspondingly renormalised as the stress increases.

\begin{figure}
\begin{center}
\includegraphics[clip,width=0.9\textwidth]{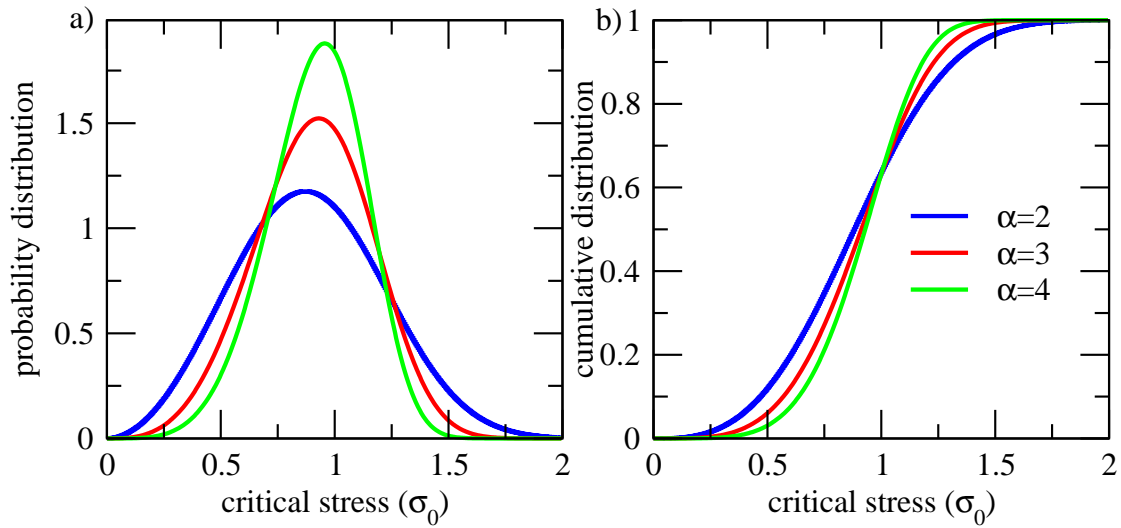}
\end{center}
\caption{Weibull distribution ($(1+\alpha)/\delta=1$ for the generalized gamma distribution, eqn.~\ref{EqProbDist}) for $\alpha=2$, $3$ and $4$ where a) plots the probability distribution function and b) plots the corresponding cumulative distribution function} \label{FigWD} 
\end{figure}

\begin{figure}
\begin{center}
\includegraphics[clip,width=0.9\textwidth]{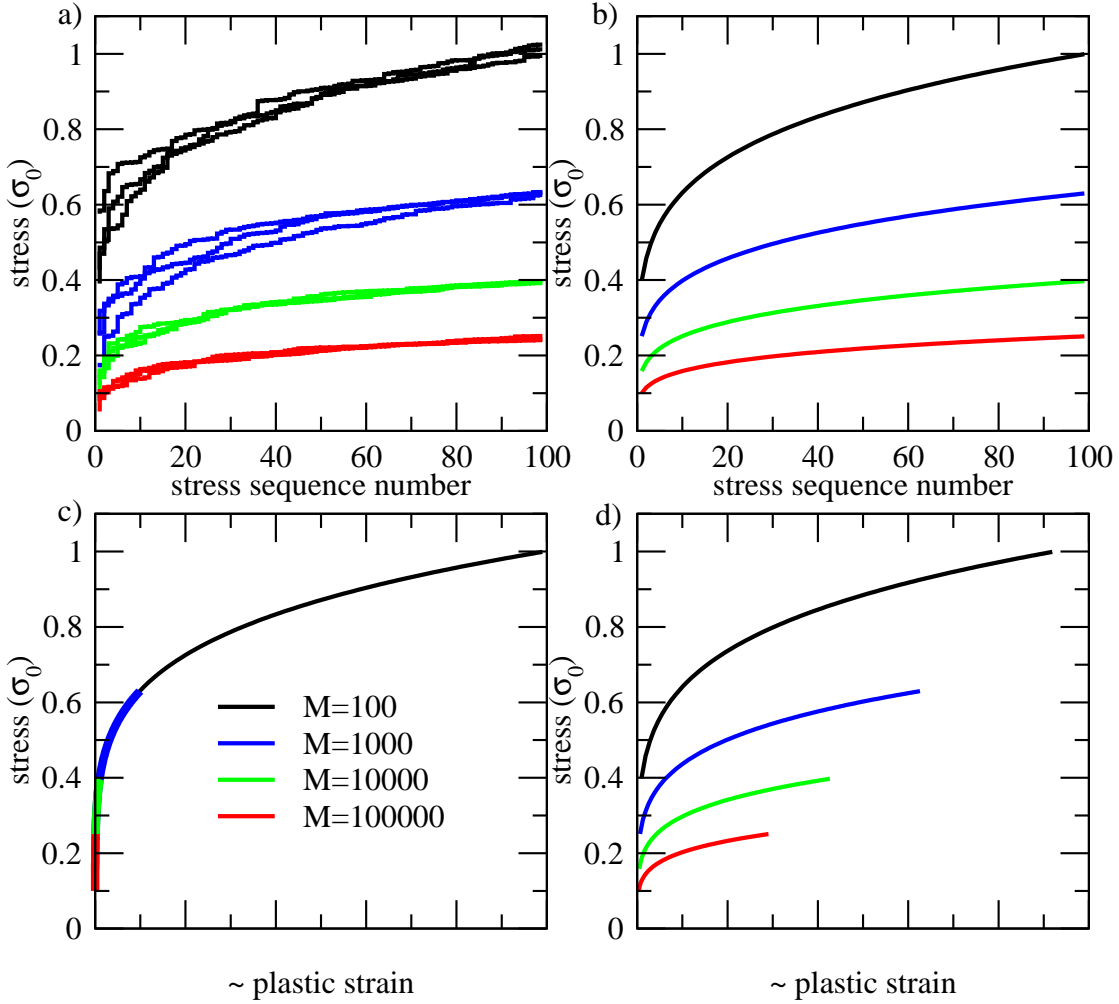}
\end{center}
\caption{a) Stochastic realizations of the ordered critical stresses derived from extreme value statistics for $M$ equal to 100, 1000, 10000 and 100000, and b) corresponding average values obtained using eqn.~\ref{EqAppB13}. Stress versus plastic strain when c) the plastic strain increment scales inversely with sample volume and d) when the plastic strain increment scales according to finite size scaling predictions of dislocation avalanche phenomena.} \label{FigData} 
\end{figure}

To generate a particular ordered list of critical stresses that would be contained in a stress-strain curve, the above procedure is iterated resulting in the following algorithm:
\begin{enumerate}
\item the applied stress, $\sigma$, is set to zero.
\item $M$ values are sampled from the critical stress distribution and sorted to produce an ordered list.
\item $\sigma$ is set equal to the lowest critical stress of the ordered list. This lowest critical stress is removed from the ordered list.
\item a new critical stress is sampled from the distribution until one is found which is larger than the current applied stress. 
\item this new critical stress is then added to the ordered list.
\item steps 3 to 5 are repeated until the desired size of the critical stress sequence is reached.
\end{enumerate}
Without loss of generality, a Weibull distribution is chosen for eqn.~\ref{EqProbDist} (i.e. $(\alpha+1)/\delta=1$) giving the probability distribution shown in fig.~\ref{FigWD}a for three values of $\alpha$. The figure shows single peaked distributions whose peak critical stresses limit to $\sigma_{0}$ for increasing $\alpha$. Fig.~\ref{FigData}a displays three stochastic realizations of stress-sequences derived using the above algorithm, spanning four orders of magnitude in $M$. Inspection of these stress sequences reveals that for low $M$ there is strong scatter in the curves indicating a high degree of stochasticity. This scatter decreases with increasing $M$, and with $M$ equal to 100000, the stress sequence almost converges to a smooth curve for all sample realisations.

In addition to the degree of stochasticity, inspection of fig.~\ref{FigData}a reveals that as $M$ decreases (reducing system size) the scale of the critical stress sequence increases. Such a size effect in stress can be rationalised via the fact that for decreasing $M$, the minimum critical stress will approach the most probable value of the distribution --- in sampling the distribution once (the extreme limit of $M=1$) the most likely value that is obtained will clearly be the most probable value ($\simeq\sigma_{0}$). On the other hand, for large $M$, the minimal critical stress will be determined largely by the extreme value properties of the distribution. This offers a quite general proposition to the statistical origin of a size effect {\em in stress} --- small sample sizes probe (on average) the mean strength of a critical stress distribution and with increasing size the (smaller) extreme value stresses of the distribution are increasingly probed. This picture forms the basis of a statistical analysis of fracture in ceramics~\cite{LawnBook} and has been used as a basis for the derivation of a size effect in stress at which the first athermal plastic event occurs~\cite{Rao2008,Parthasarathy2007,Sieradski2006,Rinaldi2007,Senger2011,Wang2012}.

An analytical expression for the {\em average} of the ordered stress list produced numerically in fig.~\ref{FigData}a will now be obtained as a function of $M$. 

The first step is to determine the first average critical stress, $\sigma_{1}^{*}$ as a function of $M$. This may be obtained via the relation~\cite{Bouchaud1997}
\begin{equation}
P_{<}[\sigma^{*}_{1}]M=1, \label{EqEVS}
\end{equation}
where $P_{<}(\sigma) $ is the cumulative distribution probability (CDF) or repartition probability:
\begin{equation}
P_{<}[\sigma]=\int_{0}^{\sigma}d\sigma'\,P[\sigma']=1-\int_{\sigma}^{\infty}d\sigma'\,P[\sigma']. \label{EqCDF}
\end{equation}
Fig.~\ref{FigWD}b shows the cumulative distribution probability (CDF) for the Weibull distribution. 

Eqn.~\ref{EqEVS} expresses the fact that there exists an average minimum stress at which the integrated number density equals unity, that is, one (minimum) critical stress exists with certainty. Clearly as $M$ increases, $\sigma_{1}^{*}$ decreases approaching zero as $M\rightarrow\infty$.

For the generalized gamma distribution,
\begin{equation}
\int_{\sigma}^{\infty}d\sigma'\,P[\sigma']=Q\left[\frac{\alpha+1}{\delta},\left(\frac{\sigma}{\sigma_{0}}\right)^{\delta}\right]
\end{equation}
where $Q[a,x]$ is the regularized incomplete gamma function with $Q[a,0]=1$. Thus eqn.~\ref{EqEVS} has the solution
\begin{equation}
\left(\frac{\sigma^{*}_{1}}{\sigma_{0}}\right)^{\delta}=Q^{-1}\left[\frac{\alpha+1}{\delta},\left(\frac{M-1}{M}\right)\right] \label{EqEVSFirstSoln}
\end{equation}
where $Q^{-1}[a,x]$ is the inverse regularized incomplete gamma function --- a function that is known.

To obtain the next mean minimum critical stress, $\sigma_{2}^{*}$ ($\sigma_{2}^{*}>\sigma_{1}^{*}$), from a sample of $M$ values, the procedure outlined in eqns.~\ref{EqEVS} to \ref{EqEVSFirstSoln} is repeated using the truncated and normalised stress distribution:
\begin{equation}
P[\sigma,\sigma^{*}]=\frac{P[\sigma]}{\int_{\sigma^{*}}^{\infty}d\sigma'\,P[\sigma']},
\end{equation}
which is valid for $\sigma>\sigma^{*}$ and zero otherwise. The corresponding truncated CDF is equal to:
\begin{equation}
P_{<}[\sigma,\sigma^{*}]=\int_{\sigma^{*}}^{\sigma}d\sigma'\,P[\sigma',\sigma^{*}]=1-\frac{\int_{\sigma}^{\infty}d\sigma'\,P[\sigma']}{\int_{\sigma^{*}}^{\infty}d\sigma'\,P[\sigma']}. \label{EqAppB5}
\end{equation}
This truncated CDF may now be used to obtain $\sigma^{*}_{2}$ via eqn.~\ref{EqEVS}: 
\begin{equation}
P_{<}[\sigma^{*}_{2},\sigma^{*}_{1}]M=\left(1-\frac{\int_{\sigma^{*}_{2}}^{\infty}d\sigma'\,P[\sigma']}{\int_{\sigma^{*}_{1}}^{\infty}d\sigma'\,P[\sigma']}\right)M=1 \label{EqAppB5a}
\end{equation}
giving
\begin{equation}
\int_{\sigma_{2}^{*}}^{\infty}d\sigma'\,P[\sigma']=\left(\frac{M-1}{M}\right)\int_{\sigma_{1}^{*}}^{\infty}d\sigma'\,P[\sigma'], \label{EqAppB6}
\end{equation}
which defines the second mean critical stress, $\sigma^{*}_{2}$, in terms of $\sigma^{*}_{1}$ and $M$.

The ordered list of mean minimum critical stresses, $\{\sigma_{1}^{*},\sigma_{2}^{*},\sigma_{3}^{*}...\}$, can now be obtained via iteration of the recurrence relation,
\begin{equation}
\int_{\sigma_{i+1}^{*}}^{\infty}d\sigma'\,P[\sigma']=\left(\frac{M-1}{M}\right)\int_{\sigma_{i}^{*}}^{\infty}d\sigma'\,P[\sigma'] \label{EqAppB9}
\end{equation}
with $\sigma_{0}^{*}=0$. Doing so gives the final result:
\begin{equation}
\int_{\sigma_{i}^{*}}^{\infty}d\sigma'\,P[\sigma']=\left(\frac{M-1}{M}\right)^{i}. \label{EqAppB10}
\end{equation}
For the generalized gamma distribution, the resulting ordered stress list is given by
\begin{equation}
\left(\frac{\sigma_{i}^{*}}{\sigma_{0}}\right)^{\delta}=Q^{-1}\left[\frac{\alpha+1}{\delta},\left(\frac{M-1}{M}\right)^i\right].\label{EqAppB13}
\end{equation}

Eqn.~\ref{EqAppB10} (and eqn.~\ref{EqAppB13} for the generalized gamma distribution) constitutes the required analytical solution to generating the mean of the ordered list of critical stresses encountered using the numerical procedure. Fig.~\ref{FigData}b displays the resulting average stress sequences for the values of $M$ used for the stochastic realizations in \ref{FigData}a for the case of the Weibull distribution, indicating good agreement between the numerical simulations and the analytical result.

Inspection of eqn.~\ref{EqAppB13} reveals that the intermittency arises from the discreteness associated with the parameter $M/(M-1)$ raised to the power of $i$ --- this is a direct result of the extreme-value-statistics approach. Those works that have also exploited the extreme-value-statistics approach, and from which a size effect in stress has been derived, have focused on the statistics of the first athermal plastic event --- the so-called weakest link regime. Eqn.~\ref{EqAppB13} embodies this result and is in fact a derivation of the first moment of the size dependent Weibull distribution used in refs~\cite{Parthasarathy2007,Sieradski2006,Rinaldi2007,Wang2012}. In these works, the initial event is then assumed to be responsible for the subsequent plasticity of the measured stress-strain curve. What is fundamentally different with the present work, is that the weakest link approach is iteratively applied to each next critical shear stress to generate the entire average shear stress sequence that would be encountered in a stress versus strain curve. That this procedure embodies extreme value statistics entails that the result is quite general and in principle will be valid for {\em all} positive valued distributions for which an inverse cumulative distribution function exists. This observation is manifested in the general expression of eqn.~\ref{EqAppB10}. Moreover the fluctuations with respect to this average are described by the Gumbel, Weibull and Fisher-Tippett universality classes of extreme value statistics~\cite{Gumbel1958,Bouchaud1997}. Thus the statistics associated with stochastic realizations of stress sequences (as in fig.~\ref{FigData}a) are themselves quite independent of the actual distribution of critical shear stresses.

The size effect evident in the stress sequence may be explicitly seen by approximating eqn.~\ref{EqAppB13} to leading order in $1/M$ (logarithmic accuracey):
\begin{equation}
\sigma_{i}\simeq\Gamma\left[1+\frac{\alpha+1}{\delta}\right]^{\frac{1}{1+\alpha}}\left(1-\left(\frac{M-1}{M}\right)^{i}\right)^{\frac{1}{1+\alpha}}\sim
\left(\frac{i}{M}\right)^{\frac{1}{1+\alpha}}\sim
\left(\frac{i}{V}\right)^{\frac{1}{1+\alpha}}, \label{EqCentralResult}
\end{equation}
where $\Gamma[a]$ is the gamma function. For logarithmic accurracy (in fig.~\ref{FigExpData}) all prefactors depending on the moments of the distribution need not be considered. 

The above result demonstrates that the size effect in stress is only influenced by the exponent $\alpha$ of the generalized gamma distribution --- an intuitive result given that the extreme value statistics regime will depend primarily on the low stress tail of the critical shear stress distribution. More generally, this implies that the leading order form, eqn.~\ref{EqCentralResult}, will be valid for a much broader class of distributions, all of which are algebraic in the limit of zero stress.

\section{Plastic strain and the emergence of a true size effect in strength} \label{SecStrain}

In a crystal, plastic strain is mediated by the sequential motion of dislocations or collections of dislocations, each one being referred to as a plastic event. Historically, in bulk crystals the individual events have been considered local when compared to the size of the material. Such a viewpoint has its theoretical origins in the early ideas of Nabarro~\cite{Nabarro1940} and Eshelby~\cite{Eshelby1957} where the corresponding far field plastic strain due to each plastic event scales inversely with sample volume. At the scale of an individual dislocation segment, this viewpoint also forms the basis of modern small strain plasticity dislocation dynamics simulations~\cite{BulatovCaiBook,ArgonBook} and a variety of coarse grained models of plastic deformation (see for example Refs.~\cite{Zaiser2006,Dahmen2009} and references therein). 

Acoustic emission experiments~\cite{Miguel2001,Weiss2003} revealed the distribution of these plastic strain magnitudes to have an algebraic component indicating scale-free physics underlies the collective motion of dislocations. Such scale free behaviour, or avalanche phenomena, indicates an underlying non-trivial complexity of dislocation based microstructure and suggests that parts of the dislocation network are in a state of self-organised criticality~\cite{Zaiser2006}. Thus plasticity belongs to a class of universal phenomenon often described as crackling noise, which encompasses such diverse phenomena as the statistics of earthquakes~\cite{Fisher1997,Dahmen1998,Mehta2006} and that of magnetic switching~\cite{Sethna2001}. Like all critical phenomenon, pure algebraic behaviour occurs only for systems without a length scale, and when a length-scale does exist the signature of approximate scale-free behaviour is how it is modified with respect to this length scale. This change in behaviour is manifested by a non-universal scaling or cut-off function which constitutes the non-algebraic part (pre-factor) of the distribution. Within this framework, the plasticity of micron sized single crystal sample volumes (investigated by both experiment~\cite{Dimiduk2006,Dimiduk2010,Zaiser2008} and dislocation dynamics simulations~\cite{Miguel2001,Ispanovity2011b,Csikor2007}) has revealed such avalanche phenomenon, where now the relevant length-scale is an external dimension. On the other hand, dislocation dynamics simulations of very long dipolar mats in which only the mobile dislocation content is explicitly modelled with the internal microstructure being fixed by a static mean-field description, also show avalanche behaviour and provides the alternative example of an internal (rather than external) length scale controlling the scaling function~\cite{Derlet2013}. In this regime of more bulk-like behaviour, where external length-scales are much larger than any internal length scale, plastic events may be again viewed as a local phenomenon with respect to sample size with their corresponding far field plastic strain scaling inversely with system size.

Thus for the bulk limit (with an internal length scale), the characteristic plastic strain magnitude, $\delta\varepsilon_{0}$, of a system with volume $V$, will be
\begin{equation}
\delta\varepsilon_{0}\sim\frac{1}{V}\sim\frac{1}{M} \label{EqEsh}
\end{equation}
giving the mean plastic strain at the $i$th critical stress $\sigma_{i}$ as
\begin{equation}
\varepsilon_{i}\simeq i\times\delta\varepsilon_{0}\sim i\times\frac{1}{M} \label{EqBulkStrain}.
\end{equation}
Eqn.~\ref{EqEsh} is a natural result of the Eshelby inclusion picture and eqn.~\ref{EqBulkStrain} exploits this fact by stating that, to logarithmic accuracy, the total plastic strain is a product of the plastic strain events and this inverse volume scaling. Again, since fig.~\ref{FigExpData} is a log-log plot, only logarithmic accuracy is needed to describe the general size effect allowing for the omission of all irrelevant prefactors in the above and in what follows.

Substitution of eqn.~\ref{EqBulkStrain} into eqn.~\ref{EqCentralResult} gives
\begin{equation}
\sigma_{i}\sim\left(\varepsilon_{i}\right)^{\frac{1}{1+\alpha}}
\end{equation}
and the result that there exists no size effect in the bulk limit. This is demonstrated in fig.~\ref{FigData}c which displays the average stress sequence now with the corresponding strain given by eqn.~\ref{EqBulkStrain} on the horizontal axis. The data corresponding to the different values of $M$ all collapse on a universal stress-plastic strain curve demonstrating that the size effect in stress is offset by a comparable size effect in plastic strain.

On the other hand, for a sample size regime, in which the scaling dimension corresponds to an external dimension, quite a different result is obtained. Both experiment and simulation reveal that the the distribution of plastic strain increments, $\delta\varepsilon$, has the asymptotic form~\cite{Zaiser2006,Csikor2007}
\begin{equation}
P(\delta\varepsilon)\sim f\left[\frac{\delta\varepsilon}{\delta\varepsilon_{\mathrm{max}}}\right]\frac{1}{\delta\varepsilon^{\tau}}, \label{EqnStrainDist}
\end{equation}
where $f[x]=\exp\left[-x^{2}\right]$ is the non-algebraic scaling function with $\delta\varepsilon_{\mathrm{max}}$ being the plastic strain scale at which the scale-free phenomenon is suppressed. The characteristic plastic strain, $\delta\varepsilon_{0}$, is given by the first moment of this distribution:
\begin{equation}
\delta\varepsilon_{0}=\frac{\int_{\delta\varepsilon_{\mathrm{min}}}^{\infty}d(\delta\varepsilon)\,\,\delta\varepsilon P(\delta\varepsilon)}{\int_{\delta\varepsilon_{\mathrm{min}}}^{\infty}d(\delta\varepsilon)\,\,P(\delta\varepsilon)}=
\delta\varepsilon_{\mathrm{max}}\frac{\Gamma\left[1-\frac{\tau}{2},\left(\frac{\delta\varepsilon_{\mathrm{max}}}{\delta\varepsilon_{\mathrm{min}}}\right)^{2}\right]}{\Gamma\left[\frac{1}{2}-\frac{\tau}{2},\left(\frac{\delta\varepsilon_{\mathrm{max}}}{\delta\varepsilon_{\mathrm{min}}}\right)^{2}\right]}\sim\delta\varepsilon_{\mathrm{max}}^{2-\tau}, \label{EqAvalanche}
\end{equation}
where $\Gamma\left[a,x\right]$ is the incomplete Gamma function. In the above $\delta\varepsilon_{\mathrm{min}}$ defines the small strain applicability limit of the distribution in eqn.~\ref{EqnStrainDist}. The last similarity in eqn.~\ref{EqAvalanche} is obtained by using the leading order approximation to the incomplete gamma function, $\Gamma\left[a,x\right]\simeq\Gamma[a]-x^{a}(1/a+\dots)$, and the knowledge that the exponent $\tau$ is approximately equal to the mean field value of $3/2$ (see sec.~\ref{SecDiscussion}).

Due to the geometry of a typical dislocation event, $\delta\varepsilon_{\mathrm{max}}$, is itself inversely proportional to a cut off length scale, $L$~\cite{Csikor2007,Zaiser2007}. A cube of volume $V=L^{3}$ gives this length scale as $L\sim M^{1/3}$ thus giving the characteristic plastic strain magnitude of the system, $\delta\varepsilon_{0}$, as
\begin{equation}
\varepsilon_{i}=i\times\delta\varepsilon_{0}\sim i\times\left(\frac{1}{M^{\frac{1}{3}}}\right)^{2-\tau}\sim i\times\left(\frac{1}{L}\right)^{2-\tau} \label{EqMicronStrain}
\end{equation}
Substitution of eqn.~\ref{EqMicronStrain} into eqn.~\ref{EqCentralResult} gives the result
\begin{equation}
\sigma_{i}\sim\left(\varepsilon_{i}\right)^{\frac{1}{1+\alpha}}\left(\frac{1}{M^{\frac{1}{3}}}\right)^{\frac{\tau+1}{\alpha+1}}\sim\left(\varepsilon_{i}\right)^{\frac{1}{1+\alpha}}\left(\frac{1}{L}\right)^{\frac{\tau+1}{\alpha+1}} \label{EqCentralCentralResult}
\end{equation}
and a true size effect in strength emerges exhibiting power-law behaviour as a function of an inverse length-scale. Fig.~\ref{FigData}d displays the corresponding stress versus plastic strain curves for the stress sequences of figs.~\ref{FigData}a-b and shows quite a distinct size effect. Note that since, $M=\rho V$, eqn.~\ref{EqCentralCentralResult} is applicable to any sample volume shape in which $V\propto L^{3}$, where $L$ is the external scaling dimension which is (only) varied.

\section{Discussion} \label{SecDiscussion}

The developed probabilistic approach results in a surprisingly simple derivation of a power law in strength with respect to sample volume. Given a distribution of critical stresses and the number $M$ of structural transformations available to a material, the two main results of the present work leading, in part, to
\begin{equation}
\sigma\sim\varepsilon^{\frac{1}{\alpha+1}}\left(\frac{1}{L}\right)^{\frac{\tau+1}{\alpha+1}}\label{EqSE}
\end{equation}
are
\begin{enumerate}
\item Intermittency in stress has its origins in the discreteness of the sequence $((M+1)/M)^{i}$. The intermittency therefore vanishes in the bulk limit of $M\rightarrow\infty$.
\item The stress of this intermittency scales inversely with $M$ and therefore sample volume.
\end{enumerate}
These two results arise directly from the extreme value statistics of the critical stress distribution, where for large enough $M$ the above universal properties emerge --- {\em independent} of the actual distribution used. Using these developments in conjunction with the known and established results of dislocation avalanche behaviour for the plastic strain distribution, results finally in eqn.~\ref{EqSE}.

What is the applicability regime of this procedure? For the extreme value statistics approach to be valid $M$ must be large enough, but need not be too large. A value of $M=100$ is already enough. Thus it is only in the limit of very small sample volumes where the present derivation is expected to break down. Indeed, as discussed in the introduction, experimentally it is known that the strength tends to saturate with system size, a regime where dislocations are largely absent and surface geometry strongly influences, via dislocation nucleation, plasticity. Thus, sample volume should be small, but not too small for the size effect to be operative. This is entirely compatible with seminal work of Uchic {\em et al} \cite{Uchic2004} who comment on the surprisingly large sample volumes ($\sim10$ microns) in which the size effect is still observed to occur.

There will also exist an upper limit in $M$ for the applicability of eqn.~\ref{EqSE}. When the sample size becomes sufficiently large, internal length scales  within the dislocation network will naturally emerge. This will reduce the importance of the external length scale and for large enough sample volumes it will dominate. In this regime, avalanche behaviour still occurs, however it is now the internal length scale which controls the finite length scale effect. From this perspective, plasticity now becomes a localized phenomenon and plastic strain will depend inversely on sample volume. Sec.~\ref{SecStrain} demonstrates that when this occurs, the size effect is absent. This limit emphasizes that the present work is entirely compatible with the change-in-mechanism approach proposed by many authors when the external dimensions of the system enter the micron regime~\cite{Kraft2010}, since the transition must ultimately manifest itself as some dislocation based mechanism. The current work does however indicate that the size effect in strength is a more general phenomenon, quite independent of any one (or few) microscopic mechanisms --- a result entirely consistent with broad applicability of fig.~\ref{FigExpData}.

\begin{figure}
\begin{center}
\includegraphics[clip,width=0.8\textwidth]{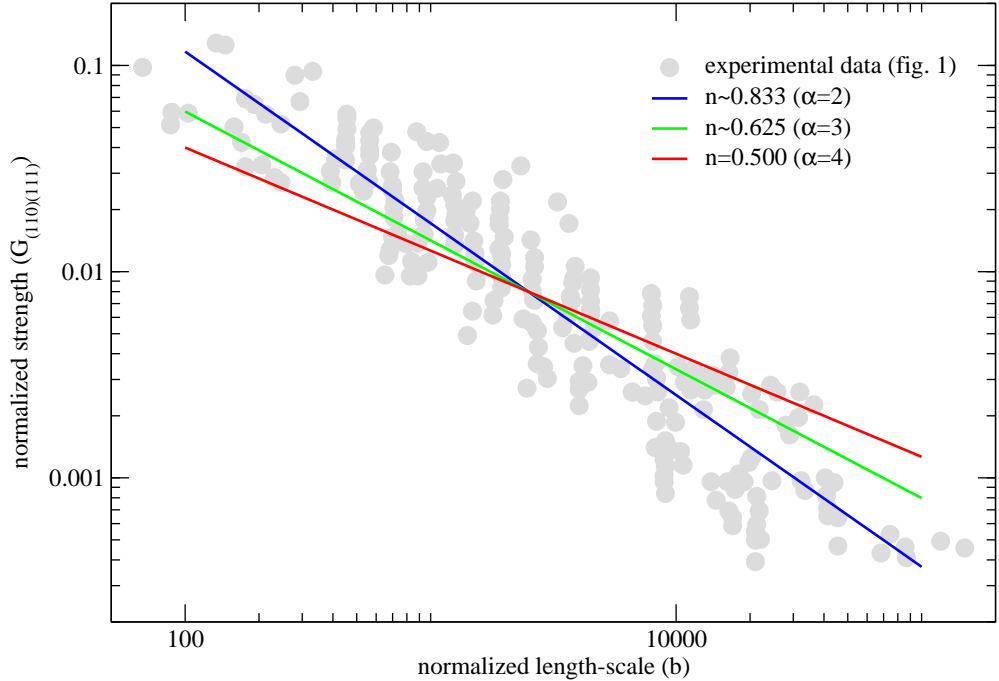}
\end{center}
\caption{Replot of experimental data in fig.~\ref{FigExpData} and size effect exponent ($n$) predictions according to eqn.~\ref{EqCentralCentralResult} ($n=(\tau+1)/(\alpha+1)$) where the mean field value $\tau=3/2$ is used, for three different values of $\alpha$ which characterizes the low stress regime of the critical shear distribution.} \label{FigFinalPlot} 
\end{figure}

The applicability of the derived model is also well reflected in various experimental reports that address the extrinsic size-effect as well as the intermittency of plastic flow with respect to internal length scales. As-prepared single crystals in the size regime of some hundred nanometres are known to fall onto the trend depicted in fig.~\ref{FigExpData}, but irradiation~\cite{Kiener2011} or the introduction of dispersed obstacles~\cite{Girault2010} in the same size range have been shown to erase the size effect in strength. In these cases dislocation-defect interaction determined by the internal length scale governs strength with the size effect now being decoupled from the external length scale of the sample. Moreover, within this regime of a dominating internal length scale, small scale mechanical testing has qualitatively shown that the strain increment magnitudes reduce when investigating similar sample sizes that contain larger pre-existing dislocation- or defect-densities~\cite{Johanns2012}. As such, the developed model is able to fully encompass some of the numerous experimentally observed stress-strain characteristics particular to some of the material systems covered in fig.~\ref{FigExpData}.

With the power-law exponent equalling, $n=(\tau+1)/(\alpha+1)$, the two free parameters of the model are the exponent of the low-critical stress end of the critical stress distribution and the scaling exponent $\tau$. Mean field calculations have demonstrated that for plastic events dominated by single slip, $\tau=3/2$~\cite{Zaiser2006,Zaiser2008}. Simulations have however shown that  this value may also be applied to multi-slip plastic events~\cite{Csikor2007}. Assuming the mean field value for $\tau$, $\alpha$ is the one free parameter of the model. Taking the values used in fig.~\ref{FigWD}a ($\alpha=2$, $3$ and $4$) gives respectively the size effect exponents $n=0.833$, $0.625$, and $0.5$. The corresponding power laws are plotted in fig.~\ref{FigFinalPlot}, along with the experimental data of fig.~\ref{FigExpData} showing good overall agreement. It should be emphasized that fitting to {\em all} of the data in fig.~\ref{FigExpData} is not a useful task, since $n$ is known to vary from material to material and also on the initial dislocation density and structure ~\cite{Shim2009,Lee2009,Schneider2013,ElAwady2013}. In the present theory this would be reflected by variations in $\alpha$ and also possibly in $\tau$. Fig.~\ref{FigFinalPlot} does however demonstrate that $\alpha$ should be larger than 2, giving a critical stress distribution that rises slowly from its zero value at zero stress.

In addition to the assumed existence of a distribution of critical shear stresses the work assumes that as plasticity evolves, the distribution becomes truncated and renormalized, but overall it retains its intrinsic form. Fig.~\ref{FigData}c puts this assumption into its appropriate context. The figure demonstrates that from the perspective of bulk plasticity, the current theory places the deformation curve of a micro-pilar experiment into the domain of micro-plasticity --- several tens of plastic events in a micro-deformation stress-strain curve will correspond to only the very early stages of bulk deformation (see for example the very early torsion experiments of Tinder and co-workers~\cite{Tinder1964,Tinder1973}). This situation is equally valid for the size affected plasticity in fig.~\ref{FigData}d, but it is somewhat hidden due to the lack of appropriate pre-factors on the strain axis. Micro-plasticity is a deformation regime where significant structural evolution and hardening are largely absent~\cite{Young1961,Vellaikal1969,Koppenaal1963}, indeed, there now exists a growing body of evidence that this is in fact the case for micro-deformation experiments~\cite{Norfleet2008,Maass2012,Oh2009,Maass2012b}. The aforementioned reference have in common that they experimentally show or suggest that there is little to no change in dislocation structure and density beyond the transition to extensive plastic flow. It is this perspective that gives justification to the assumption of an unchanging (but continuously truncated) critical stress distribution. Put in other words, it is recognized that upon a discrete plastic event occurring the internal structure of the sample volume changes non-negligibly, however the next critical shear stress which characterises this new configuration, is still drawn from the same critical stress distribution since the characteristic internal length scale and dislocation density can not change in any significant way due to this event occurring --- this is the essence of the micro-plastic regime. Finally, the assumption that $M$ does not change with plastic evolution turns out not to be necessary for the derivation of eqn.~\ref{EqSE}. Indeed $M$ could either decrease or fluctuate around some mean value with respect to the plastic evolution and the same scaling result would be obtained.

\section{Concluding remarks}

In summary, the size effect paradigm ``smaller is stronger'', as embodied by the power law, $d^{-n}$, is shown to originate from a combination of a size effect in stress derived from the extreme value statistics of an assumed distribution of critical stresses, and a size effect in strain derived from the finite scaling associated with scale-free dislocation activity. Both contributions may be considered universal, depending little on the fine details of a particular material. In particular, $n=(\tau+1)/(\alpha+1)$, where $\alpha$ is the leading order algebraic exponent of the low stress regime of the critical stress distribution and $\tau$ is the scaling exponent associated with the distribution of plastic strain magnitudes.

\end{document}